# Fermiological Interpretation of Superconductivity/Non-superconductivity of FeTe$_{1-x}$Se$_x$ Thin Crystal Determined by Quantum Oscillation Measurement

H. Okazaki,[1] T. Watanabe,[1,2] T. Yamaguchi,[1] K. Deguchi,[1,2] S. Demura,[1,2] S. J. Denholme,[1] T. Ozaki,[1] Y. Mizuguchi,[3] H. Takeya,[1] T. Oguchi,[4] Y. Takano[1]

[1]*National Institute for Materials Science, 1-2-1 Sengen, Tsukuba, Ibaraki 305-0047, Japan*

[2]*Graduate School of Pure and Applied Sciences, University of Tsukuba, 1-1-1 Tennodai, Tsukuba, Ibaraki 305-8577 Japan*

[3]*Department of Electrical and Electronic Engineering, Tokyo Metropolitan University,1-1, Minami-osawa, Hachioji, 192-0397, Japan*

[4] *Institute of Scientific and Industrial Research, Osaka University, 8-1 Mihogaoka, Ibaraki, Osaka 567-0047, Japan*

E-mail : OKAZAKI.Hiroyuki@nims.go.jp

**Abstract**

We have successfully observed quantum oscillation (QO) for FeTe$_{1-x}$Se$_x$. QO measurements were performed using non-superconducting and superconducting thin crystals of FeTe$_{0.65}$Se$_{0.35}$ fabricated by the scotch-tape method. We show that the Fermi surfaces (FS) of the non-superconducting crystal are in good agreement with the rigid band shift model based on electron doping by excess Fe while that of the superconducting crystal is in good agreement with the calculated FS with no shift. From the FS comparison of both crystals, we demonstrate the change of the cross-sectional area of the FS, suggesting that the suppression of the FS nesting with the vector $\boldsymbol{Q}_s = (\pi, \pi)$ due to excess Fe results in the disappearance of the superconductivity.

## 1. Introduction

Fe-based superconductors have stimulated fundamental discussions on the mechanism of superconductivity. In these discussions, knowing the nature of elementary excitations and interplay between Fermi surfaces (FS) is essential to understand the origin of its superconductivity. It has been suggested that antiferromagnetic (AF) spin fluctuation due to the interband correlation in doped Fe-base compounds is important to understand the mechanism of superconductivity [1,2] since parent Fe-based compounds exhibit AF ordering associated with the FS nesting connected by the wave vector $\boldsymbol{Q}_s = (\pi, \pi)$. On the other hand, other calculations predicted that the orbital fluctuation generates the pairing interaction in Fe-based superconductors [3]. Thus, it is very important to clarify the FS experimentally. Among Fe-based superconductors, 11 type superconductors have been extensively studied in order to understand the mechanism of Fe-based superconductivity since 11 type compounds have the simplest structure composed of only superconducting layers. Additionally, in the 11 type superconductor, iron-selenide compounds exhibit a high superconducting transition temperature ($T_c$). The $T_c$ of FeSe ($T_c = 8$ K) can be increased up to about 40 K by applying pressure or by intercalation of an alkali metal or molecules between FeSe layers [4-7]. Therefore, the information of its FS is important to elucidate Fe-based superconductivity and may shed light on the achievement of a higher $T_c$. Quantum oscillation measurements are a powerful technique to observe the FS, but has not been reported for the 11-type compounds. It can be considered that a single crystal of 11 type compound is not homogeneous as the QO can be observed.

Recently, we fabricated several thin crystals of $FeTe_{1-x}S_x$ by the scotch-tape method, which is a technique to easily take an ultrathin crystal from a bulk material such as graphene [8], and we observed the different temperature dependence of resistivities of each crystal [9]. Additionally, we observed a crystal with high superconducting performance compared that of the bulk. The results present that homogeneous crystals can be taken from the bulk by the scotch-tape method. If a homogeneous section can be picked out from a 11 type compound by the scotch-tape method, the quantum oscillation is probably observed. We measured the magnetic field dependence of resistivity of the FeTe$_{1-x}$S$_x$ thin crystals fabricated by the scotch-tape method, and successfully observed the Shubnikov-de Haas oscillation.

2. Experimental

Single crystals of FeTe$_{0.65}$Se$_{0.35}$ are prepared by the self-flux method. The powder of Fe (99.9%), Te (99.999%), and Se (99.9999%) were mixed in the appropriate ratios and sealed into an evacuated quartz ampoule. The sealed materials were heated at 1100 °C for 20 hours and then cooled in a temperature gradient of 2 °C/h down to 650 °C followed by furnace cooling. The obtained single crystals were placed on the scotch tape, and cleaved several times. The scotch tape with the FeTe$_{1-x}$Se$_x$ flakes were attached to an oxidized silicon substrate, and pressed to enhance the bonding between the crystals and the substrate by van-der-Waals attraction. By this process, thin crystals were left on the substrate.

We prepared the four-terminal electrodes by electron-beam lithography because the area of thin crystals is small. Both the width and separation of the terminals are 1 μm. Resistivity measurements were performed from 0 to 15 T using a superconducting magnet. The electrical current and magnetic field were applied parallel to the *ab* plane and *c* axis, respectively.

Band-structure calculations for FeSe are performed by the first-principles full-potential linearized augmented plane-wave method within the generalized gradient approximation. Methodological details follow those used in the previous calculation [10].

3. Results and discussion

Figure 1 shows resistivities of FeTe$_{0.65}$Se$_{0.35}$ thin crystals as a function of temperature *T*. The thin crystal with thickness of 30 nm exhibits semiconductivity without superconductivity while the thin crystal with thickness of 100 nm clearly shows a superconducting transition. The difference can be explained by the different concentration of excess Fe since the excess Fe results in weak charge carrier localization and the change from metallic to semiconducting behavior above $T_c$ [9,11,12]. The non-superconducting thin crystal has certain excess Fe and the superconducting thin crystal has a low amount of excess Fe.

Figure 2(a) shows the magnetic field B dependence of resistivity in non-superconducting thin crystal for $B//c$ axis at 1.8 K. The oscillations are clearly observed in the non-superconducting sample after subtracting the smooth background, as shown in Fig 2(b). Thus, we found that the quantum oscillation can be observed in the 11 type compound since a homogeneous crystal is extracted from the single crystal bulk by the scotch-tape method. Fourier transforms of the SdH oscillation is shown in Fig. 2(c). We found three fundamental frequencies $F_\alpha$ = 320 T, $F_\beta$ = 160 T, and $F_\beta$ = 45 T which we shall call the $\alpha$, $\beta$ and $\gamma$, respectively. The peaks around 90 and 120 T are the second and third harmonic of $\gamma$. The fundamental frequency $F$ is proportional to the k-space cross sectional area $A_k$ of Fermi surface (FS): $F = (\hbar/2\pi e) A_k$. Applying simple approximation $A_k = \pi k_F^2$, where $k_F$ is the radius of the FS, $k_F^\alpha$, $k_F^\beta$, and $k_F^\gamma$ were estimated to be 0.099, 0.070, and 0.037 Å$^{-1}$, respectively.

We also observed the quantum oscillation for the superconducting crystal. Generally, it is difficult to observe the quantum oscillation of a superconductor because of the zero resistivity below $T_c$ and the thermal broadening of the Landau levels above $T_c$. Thus, we measured the magnetic field B dependence of resistivity using the current above the critical current at 1.9K, as shown in Fig. 3. From the same analysis, we found a fundamental frequency $F_\delta$ = 220 T denoted by $\delta$, corresponding to $k_F^\delta$ = 0.082 Å$^{-1}$.

It is essential to compare the FS of both crystals because one crystal shows the superconductivity while the other does not. Figure 4(a) and (b) show the Fermi surfaces based on the estimated $k_F$, and comparison with theoretical results. The obtained FS for the non-superconducting crystal are inconsistent with the calculated FS of FeSe. Since the crystal does not exhibit the superconductivity, it is necessary to consider the effect of excess Fe which suppresses the superconductivity. Considering that an excess Fe provides the Fe-chalcogenide layers with two electrons, we compared the FS for the non-superconducting crystal with FS of rigid band shift model based on electron doping due to excess Fe. The rigid band shift=results in the change of the FS area. The obtained FS of the non-superconducting crystal are in good agreement with the rigid band shift of +0.15 eV, as shown in Fig. 4(a). We found that the obtained FS of $k_F^\beta$ and $k_F^\gamma$ are in agreement with the hole pockets at Γ point in the Brillouin zone and the area of α-FS is close to that of small electron pocket at M point. With rigid band shift toward positive energy, the hole pockets at Γ point shrink while the electron pockets at M point enlarge. For rigid band shift of +0.15 eV, the hole pockets are smaller than the electron pockets. Thus, the FS for the non-superconducting crystal is probably the electronic structure difficult to occur FS nesting with the vector $Q_s$ which is considered as the driving force of Fe-based superconductivity [1-3]. Electron doping due to excess Fe, corresponding to rigid band shift toward positive energy, results in the suppression of FS nesting. In contrast, the FS of the superconducting crystal is in good agreement with the calculated FS of FeSe, as shown in Fig. 4(b). The FS estimated by $k_F^\delta$ is in good agreement with both small hole pocket at Γ and small electron pocket at M in the comparison of FS with calculated FeSe with no rigid band shift model. This result implies that similar FS, which are responsible for the nesting vector $Q_s$, exist at Γ and M point.

We also estimated the amount of excess Fe from the value of rigid band shift and the density of state, as shown in right panel of Fig. 4(c). The amount is estimated to be 0.07 for the non-superconducting crystal, indicating that the composition the non-superconducting crystal is $Fe_{1.07}Te_{0.65}Se_{0.35}$. On the other hand, the amount of superconducting crystal is $Fe_{1.00}Te_{0.65}Se_{0.35}$ since the superconducting crystal can be explained by no rigid band shift. Therefore, we expect that the non-superconductivity for $Fe_{1.07}Te_{0.65}Se_{0.35}$ crystal is probably attributed to the suppression of FS nesting due to the change of FS area and $Fe_{1.00}Te_{0.65}Se_{0.35}$ crystal shows the superconductivity with the nesting vector $Q_s$. Our results suggest that the suppression of the FS nesting by the excess Fe suppresses the superconductivity.

## 4. Conclusion

We successfully observed the Shubnikov-de Haas oscillation in FeTe$_{0.65}$Se$_{0.35}$ thin crystals with non-superconductivity and superconductivity fabricated by the scotch-tape method. From the Fourier transforms of the Shubnikov-de Haas oscillation, the Fermi surfaces (FS) of both crystals were identified and compared with the calculated FS of 11 type compound FeSe. Experimental FS of the non-superconducting crystal are in good agreement with the rigid band shift model based on electron doping due to excess Fe while that of the superconducting crystal is consistent with no rigid band shift model. From the rigid band shift, we estimated the amount of excess Fe and demonstrated the influence of excess Fe on the FS. The comparison of the FS of non-superconducting Fe$_{1.07}$Te$_{0.65}$Se$_{0.35}$ with superconducting Fe$_{1.00}$Te$_{0.65}$Se$_{0.35}$ suggests that excess Fe suppresses the suppression of nesting vector $Q_s$ and results in the absence of the superconductivity is disappeared.

**Acknowledgements**

This work was supported partly by JST-TRiP and JST-EU-Japan SC.

**Figure captions**

Fig. 1.   Temperature dependence of resistivity of FeTe$_{0.65}$Se$_{0.35}$ thin crystals with the thickness of 30 and 100 nm.

Fig. 2.   (a) Magnetic field dependence of resistivity of non-superconducting crystal with the thickness of 30 nm and non-superconductivity at 1.8 K. (b) SdH oscillation extracted from the magnetic field dependence of resistivity in (a). (c) Fourier transform of the SdH oscillation.

Fig. 3.   (a) Magnetic field dependence of resistivity of superconducting crystal with the thickness of 100 nm and superconductivity ($T_c$ = 15.0 K) at 1.9 K. (b) SdH oscillation extracted from the magnetic field dependence of resistivity in (a). (c) Fourier transform of the SdH oscillation.

Fig. 4. Comparison of FS derived from SdH oscillation with the calculated FS of FeSe. Broken lines and solid lines represent experimental and calculated FS, respectively. In the calculation, the FS consists of hole pockets at $\Gamma$ and electron pockets at M point. (a) FS of non-superconducting crystal and rigid band shift model of +0.15 eV. (b) FS of superconducting crystal and no rigid band shift. (c) Calculated band dispersion and the DOS of FeSe.

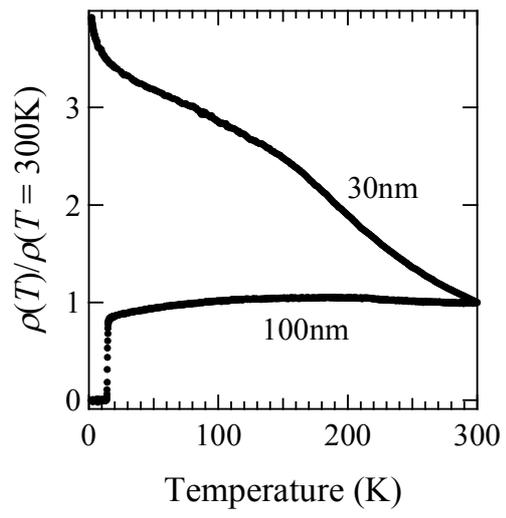

Fig. 1

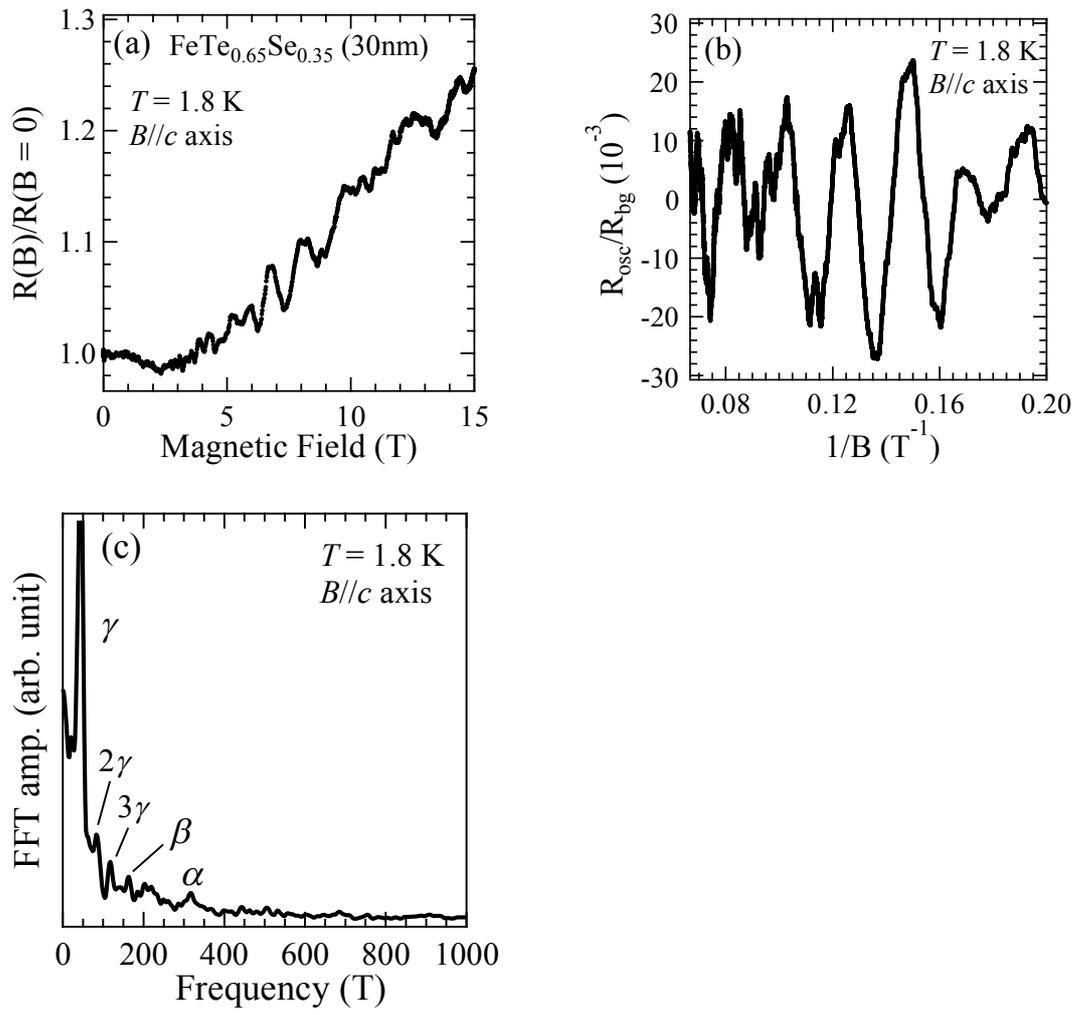

Fig. 2

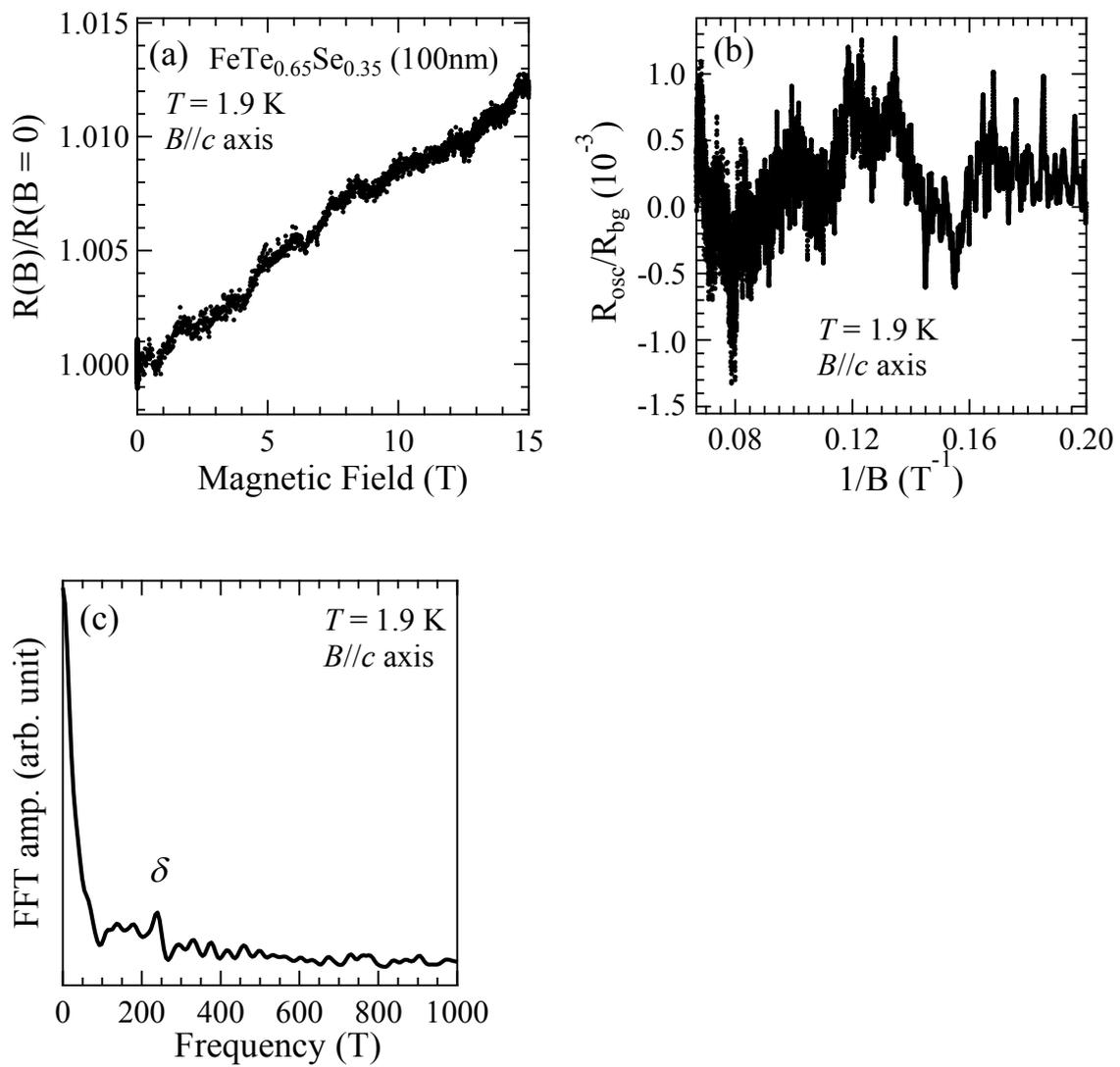

Fig. 3

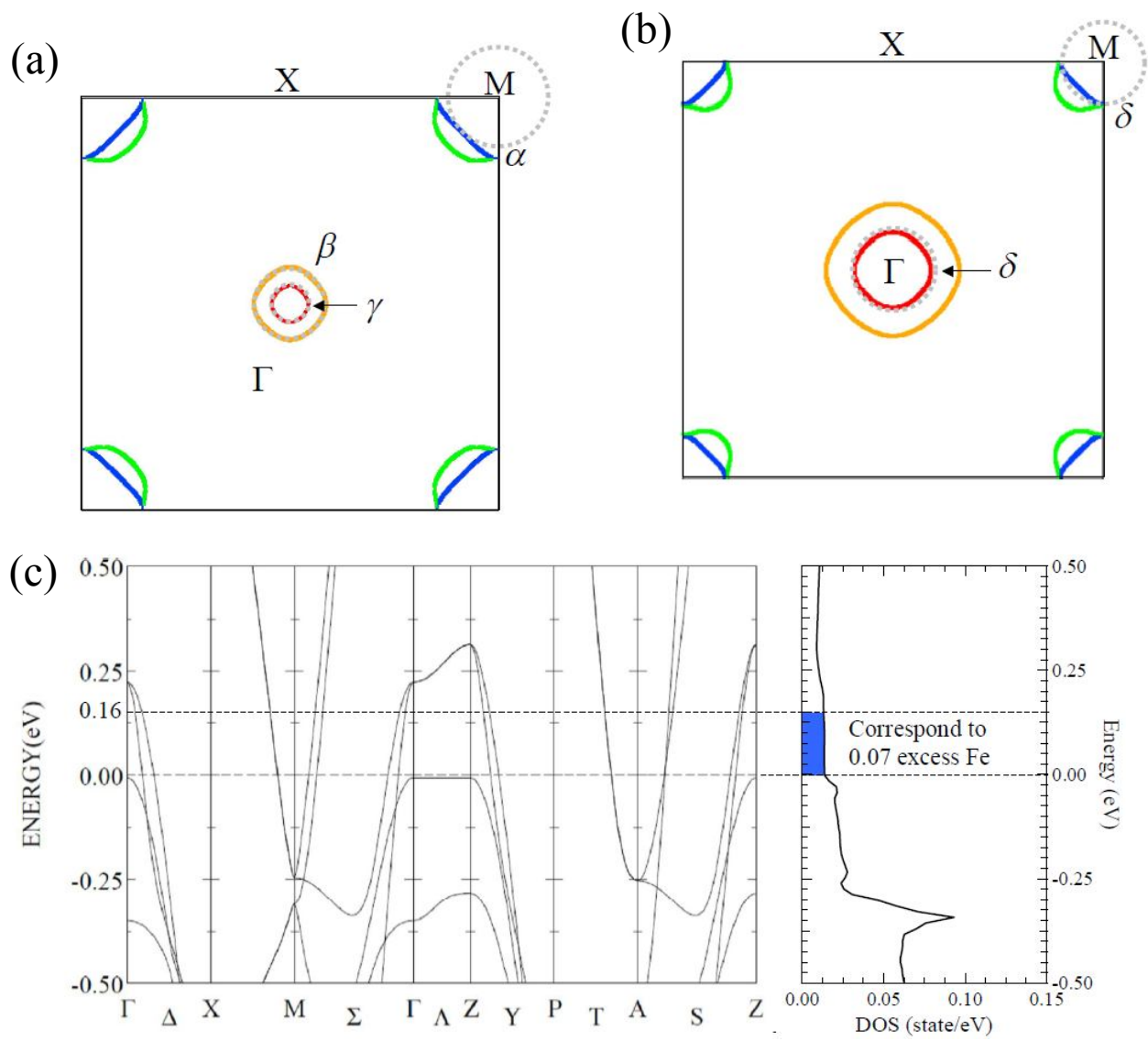

Fig. 4